\documentclass[onecolumn,numberedappendix]{openjournal}

\pdfoutput=1
\makeatletter
\def\frontmatter@title@below{\vspace*{-2.63\baselineskip}\vspace*{0.25in}}%
\makeatother
\usepackage[T1]{fontenc}
\usepackage[utf8]{inputenc}

\usepackage[section]{placeins}
\usepackage{xcolor}
\usepackage{dsfont}
\usepackage{bm}
\usepackage{mathtools}
\usepackage{enumerate}
\usepackage{natbib}
\usepackage{hyperref}
\definecolor{linkcolor}{rgb}{0.0,0.3,0.5}

\hypersetup{
    unicode,
    colorlinks=true,
    linkcolor=linkcolor,
    citecolor=linkcolor,
    filecolor=linkcolor,
    urlcolor=linkcolor
}
\shorttitle{Cosmological peculiar velocities in general relativity}
\shortauthors{Clarkson \& Maartens}
\submitted{Version \today}
\journalinfo{Open Journal of Astrophysics}

\def\be{\begin{equation}}
\def\ee{\end{equation}}
\def\bea{\begin{eqnarray}}
\def\eea{\end{eqnarray}}

\newcommand{\sd}{\mathrm{d}}
\newcommand{\ud}{\mathrm{D}}

\definecolor{MyB}{rgb}{0.1,0.1,1.0}

\newcommand{\tsagas}{Tsagas et al.}

\begin{document}

\title{Cosmological peculiar velocities in general relativity}

\author{Chris Clarkson}
\affiliation{Centre for Theoretical Physics \& Astronomy, 
Queen Mary University of London, London E1 4NS, UK}
\affiliation{Department of Physics \& Astronomy, University of the Western Cape, Cape Town 7535, South Africa}
\email{chris.clarkson@qmul.ac.uk}

\author{Roy Maartens}
\affiliation{Department of Physics \& Astronomy, University of the Western Cape, Cape Town 7535, South Africa}
\affiliation{National Institute for Theoretical \& Computational Sciences, Cape Town 7535, South Africa}
\email{rmaartens@uwc.ac.za}

\begin{abstract}
We reconsider the late-time evolution of galaxy peculiar velocities in the
$1+3$ covariant approach to cosmological perturbation theory. It has recently
been claimed that this approach predicts substantially stronger growth of
peculiar velocities than standard metric-based perturbation theory -- on the
grounds that the covariant treatment is fully relativistic whereas standard
treatments are effectively Newtonian. We show that this is not the case. When
the covariant equations are applied consistently, the $1+3$ approach reproduces
exactly the standard perturbative result for peculiar-velocity growth. The
stronger growth laws claimed in recent work arise from an inconsistent
treatment of the coupled covariant system, in which terms constrained by the
field equations are treated as if they were independent sources. Further claims
are made that the stronger bulk flows can mimic accelerated expansion in a dust
universe. We argue that these claims rest on a confusion between the kinematics
of an arbitrarily chosen observer congruence and the physical expansion of the
matter congruence traced by galaxies. We conclude that the standard treatment
of peculiar velocities is correct and fully relativistic~-- and does not lead
to anomalous bulk flows or to apparent accelerated expansion.
\end{abstract}


\section{Introduction}

Large-scale peculiar velocities provide an important test of cosmological structure formation and of the relativistic description of inhomogeneities. Recent work \citep{Tsaprazi:2019bbi,Maglara:2022pmz,Tsagas:2024zrx,Tsagas:2025nic,Tsagas:2025pxi,Pasten:2026fwd} (hereafter \tsagas) and related earlier papers, e.g. 
\citet{Filippou:2020gnr,Tsagas:2021dsl,Tsagas:2021tqa,Asvesta:2022fts},
has argued that the standard analysis of peculiar velocities in a matter-dominated universe is effectively Newtonian, and that a fully relativistic covariant analysis predicts substantially stronger growth. In this picture, peculiar velocities are defined relative to a chosen `cosmic rest frame' whose observers move with respect to the matter. There is nothing wrong in introducing  a reference congruence as a mathematical tool~-- in metric-based perturbation theory this is just part of a gauge choice (see e.g. \cite{Malik:2008im}). The issue is the claim in \tsagas\  that the $1+3$ covariant approach \citep{ehlers1993contributions,ellis:2009} yields a stronger growing mode, with
\begin{align}\label{sjdkbncsdkcn}
v^a \propto t^{4/3}\,,
\end{align}
in perturbed Einstein-de Sitter, rather than the standard result in longitudinal gauge,
\begin{align}
v^a \propto t^{1/3} \,.
\end{align}
The rapid growth rate~\eqref{sjdkbncsdkcn} has been used in \tsagas\ to argue that bulk motions can in fact mimic accelerated expansion in a dust universe~\citep{Sah:2024csa}. (As in \tsagas\ we ignore a cosmological constant in this work.)

If correct, this would obviously have major implications.   If peculiar velocities at decoupling are of order $10\,{\rm km\,s^{-1}}$, then the standard EdS growth $v\propto a^{1/2}$ gives present-day velocities of order a few $10^2\,{\rm km\,s^{-1}}$, broadly as expected and measured. By contrast, the growth law $v\propto t^{4/3}\propto a^2$ amplifies velocities by a factor $(1+z_{\rm dec})^2\sim10^6$, leading today to $v\sim10^7\,{\rm km\,s^{-1}}$, i.e. tens of times the speed of light. Even the weaker claim $v\propto t\propto a^{3/2}$ gives an amplification of order $(1+z_{\rm dec})^{3/2}\sim 4\times10^4$, implying present-day velocities of order $4\times10^5\,{\rm km\,s^{-1}}$, already relativistic, and orders-of-magnitudes larger than measured in peculiar velocity fields.

We show here that these claims do not follow from the covariant equations. When the system is treated consistently, the covariant approach reproduces the standard relativistic result.
The main points are as follows.
\begin{itemize}
\item The claimed growth law $v^a\propto t^{4/3}$ is spurious. Once all relevant general relativistic equations are imposed, the covariant system gives the standard EdS result
\begin{align}
v^a\propto t^{1/3}\propto a^{1/2}\,,
\end{align}
in agreement with earlier covariant analyses \cite{Maartens:1998qw,Ellis:2001ms} and with standard metric-based perturbation theory.

\item The standard result is not a Newtonian approximation. It is the outcome of linear relativistic perturbation theory in a particular gauge, and the corresponding covariant treatment is fully equivalent at first order.

\item The stronger growth claimed in \tsagas\ arises from an inconsistent treatment of the coupled covariant system, in which terms constrained by the field equations are treated as if they were independent source terms.

\item The associated claim that bulk flows can generate the illusion of acceleration in a matter-dominated universe rests on a confusion between the kinematics of an abstract mathematical observer congruence and the physical expansion of the matter congruence traced by galaxies.

\item Fully relativistic numerical simulations likewise do not support any large enhancement of scalar peculiar-velocity growth over the standard result \citep{Adamek:2015eda,Adamek:2017grt}; relativistic corrections are small and arise from effects beyond those invoked in \tsagas.

\end{itemize}

The first issue to check is whether the disagreement could simply be due to a different choice of reference frame, since peculiar velocities are frame-dependent~-- clearly one can define a mathematical reference frame with any peculiar velocity growth rate. In fact, for scalar perturbations of EdS (the models considered in \tsagas), the shear-free and irrotational reference congruence used in the covariant analysis is just the covariant counterpart of the longitudinal-gauge observer congruence, as we show in Appendix~\ref{appb}. The disagreement is therefore real.

The origin of the problem is easy to identify. Equations such as Eq.~(15) of \cite{Maglara:2022pmz} take the form
\bea \label{eq7}
\ddot v_a+\frac{1}{2}H \dot v_a - 2 H^2 v_a =
\mathcal{S}_a\,,
\eea
where $\mathcal{S}_a$ contains density and velocity perturbations. The error is not in \eqref{eq7} itself, but in treating $\mathcal{S}_a$ as if it were an external source that can be neglected in order to isolate a physically relevant homogeneous solution. Setting $\mathcal{S}_a=0$ as in \tsagas\ to compute the `homogeneous solution' leads to
\begin{align}\label{eq7a}
\ddot v_a+\frac{1}{2}H \dot v_a - 2 H^2 v_a =0
\quad\Longrightarrow\quad
v_a\propto t^{4/3}\,.
\end{align}
This step is illegitimate. The quantity $\mathcal{S}_a$ is not independent of $v_a$: it is fixed by the remaining covariant field equations and is not an external driving term in~\eqref{eq7}, which is required when solving for a homogeneous solution to a forced differential equation. More generally, different but equivalent rearrangements of the same coupled system lead to different second-order equations for $v_a$, so the split into `homogeneous' and `source' parts has no invariant physical meaning. In Sec. \ref{secpec} we show that, once the coupled system is handled consistently, the covariant analysis reduces to the standard result $v^a\propto t^{1/3}$.

Two related confusions underlie the disagreement. The shear-free,
irrotational reference congruence is identified in \tsagas\ with a Newtonian
approximation, when in fact it is the covariant counterpart of longitudinal
gauge: a fully relativistic solution of Einstein's equations with a
particular frame/gauge fixed, not a physical limit, or approximation beyond the perturbation theory. Conversely, the
additional kinematic freedom that appears when this choice is relaxed~--
we show below that the shear and 4-acceleration of an auxiliary congruence can  be growing unphysically~-- is treated as
new relativistic physical content. 
The point is not contradicted by the Stewart-Walker lemma~\citep{Stewart:1974uz}. Once a congruence is
chosen, quantities such as $A_a$ and $\sigma_{ab}$ vanish in the FLRW background
and are gauge-invariant first-order variables. But they are still variables of
the chosen congruence. Changing the congruence changes them, so they do not by
themselves represent new dynamical degrees of freedom of the dust.
This is starkest in the $E_{ab}=0$ case: the spacetime is
then exact FLRW~-- there is no physical perturbation at all~-- yet $v_a$,
$A_a$ and $\sigma_{ab}$ can take essentially any time dependence one
chooses, describing a non-trivially tilted observer field in an unperturbed
background. Without a physical force to drive the acceleration of that
observer field, any associated growth law for $v_a$ is irrelevant to the
actual dynamics.

A separate claim in \tsagas\ is that the larger bulk velocities inferred from \eqref{eq7a} can produce the illusion of accelerated expansion in a matter-dominated universe, removing the need for dark energy. However,  this separate claim further relies on conflating two different congruences: the matter worldlines, traced by galaxies, and the arbitrarily chosen observer worldlines that define the mathematical cosmic rest frame. The physical expansion and deceleration of the universe are properties of the matter congruence: they do not depend on the choice of reference congruence used to describe peculiar motions. Observers moving with respect to the matter may describe the same spacetime differently, but after applying the appropriate relativistic transformations they still have to observe the same decelerating matter expansion~-- this is a simple application of general covariance. 

Ultimately the source of confusion concerns the role of the 4-acceleration and momentum flux of the chosen observer congruence. These are  presented as genuinely relativistic drivers of enhanced peculiar-velocity growth. On the contrary, they simply reflect the fact that the reference congruence is not comoving with the dust. The matter itself remains geodesic on the scales of interest, as appropriate for pressure-free dust, and the `observer' 4-acceleration and associated flux have no physical effect on the expansion of galaxy worldlines.

\section{Covariant cosmology and peculiar velocities}
\label{seccov}

We follow \tsagas\ and use the $1+3$ covariant approach. For the convenience of readers not familiar with this formalism, we summarise here the elements needed below. More detailed accounts may be found in \cite{ehlers1993contributions,ellis:2009,Ellis:1998ct,Tsagas:2007yx,Ellis:2012}.

\subsection{Basic definitions}

The starting point is a choice of timelike 4-velocity $u^a$, with $u_a u^a=-1$. This defines the projection tensor into the rest-space orthogonal to $u^a$:
$
h_{ab}=g_{ab}+u_a u_b\,.
$
The covariant derivative of $u^a$ is irreducibly decomposed as
\bea
\label{expu}
\nabla_b u_a = \frac{1}{3}\Theta\, h_{ab}+\sigma_{ab}+\omega_{ab}-u_bA_a\,,
\eea
where $A_a=u^b\nabla_bu_a$ is the 4-acceleration, $\Theta=\nabla_a u^a$ is the volume expansion rate, $\sigma_{ab}=h_{\langle a}^{\ \ c}h_{b\rangle}^{\ \ d}\nabla_d u_c$ is the shear tensor, and $\omega_{ab}=h_{[a}^{\ \ c}h_{b]}^{\ \ d}\nabla_d u_c$ is the vorticity tensor. Thus the kinematics of any congruence are described by its expansion, distortion, rotation and acceleration.

Angle brackets denote the projected, symmetric, trace-free part of a tensor:
$
J_{\langle ab\rangle}
= h_{(a}^{\ \ c}h_{b)}^{\ \ d}J_{cd}
-\frac13 h_{ab}h^{cd}J_{cd}\,.
$
The covariant time derivative along $u^a$ is
$
\dot J_{a\cdots}^{\ \ \ \cdots b}=u^c\nabla_c J_{a\cdots}^{\ \ \ \cdots b}\,,
$
while the projected spatial derivative is
$
\ud_c J_{a\cdots}^{\ \ \ \cdots b}
= h_a^{\ d}\cdots h_e^{\ b} h_c^{\ f}\nabla_f J_{d\cdots}^{\ \ \ \cdots e}\,.
$

In exact FLRW spacetime, when $u^a$ is normal to the homogeneous spatial hypersurfaces,
\begin{align}
\Theta=3H\,,\qquad A_a=0\,,\qquad \sigma_{ab}=0\,,\qquad \omega_{ab}=0\,,
\end{align}
where $H$ is the {standard} Hubble rate.

\subsection{Relations between frames}
\label{secfram}

A second 4-velocity $\tilde u^a$ is related to $u^a$ by a relative spacelike velocity field $v^a$:
\begin{align} \label{tuu}
\tilde u^a=\gamma(u^a+v^a)\,,
\qquad
u_a v^a=0\,,
\qquad
\gamma=(1-v_a v^a)^{-1/2}\,.
\end{align}
Each 4-velocity defines a congruence of worldlines and hence a frame of observers. In general a frame is not physically relevant unless it is tied to some matter content such as the radiation which defines the CMB frame, for example, though it may be of mathematical convenience.  In the late-time cosmological context, following \tsagas, we identify $\tilde u^a$ as the matter 4-velocity, i.e. the 4-velocity of observers comoving with the galaxies. The matter is taken to be pressure-free dust in free fall. The field $u^a$ then defines a chosen `cosmic rest frame' whose `observers' move with respect to the matter. This rest frame will turn out to be the coordinate frame of the longitudinal (Newtonian) gauge (see Appendix~\ref{appb}).

Assuming $|v^a|\ll1$, we neglect terms nonlinear in $v^a$ and its derivatives. The exact relations \citep{Maartens:1998xg,Maartens:1998qw} between the kinematic quantities of the two congruences then reduce to
\bea \label{tkin}
\tilde\Theta &=& \Theta+\ud_a v^a + A_a v^a\,,\\
\tilde A_a &=& A_a+h_a^{\ b}\dot v_b+\frac13\Theta v_a+\sigma_{ab}v^b+\omega_{ab}v^b + A_b v^b u_a\,,\\
\tilde\sigma_{ab} &=& \sigma_{ab}+\ud_{\langle a}v_{b\rangle}+A_{\langle a}v_{b\rangle}+2u_{(a}\sigma_{b)c}v^c\,,\\
\tilde\omega_{ab} &=& \omega_{ab}-\ud_{[a}v_{b]}+A_{[a}v_{b]}-2u_{[a}\omega_{b]c}v^c\,.
\eea
These relations may be inverted by removing the tildes on the left-hand side, placing tildes on the right-hand side, and replacing $v^a\to -v^a$.

The important point is that these are relations between the kinematic quantities of two distinct congruences. Consequently, to relate the two sets of kinematic quantities one needs not only $v_a$ but also $\nabla_b v_a$ at each event. This reflects the freedom in the choice of cosmic rest frame.

\subsection{Matter variables}

The matter energy-momentum tensor is a physical tensor, independent of the observer congruence. In the matter frame it takes the simple dust form
\begin{align}
T_{ab}=\tilde\rho\,\tilde u_a\tilde u_b\,.
\label{emt}
\end{align}
Equivalently, relative to the $u^a$ frame,
\begin{align}
T_{ab}=\rho\,u_a u_b + p h_{ab}+2q_{(a}u_{b)}+\pi_{ab}\,,
\end{align}
where $\rho$, $p$, $q_a$ and $\pi_{ab}$ are the density, pressure, momentum flux and anisotropic stress measured by the $u^a$ observers.

To first order in $v^a$, for pressure-free dust, $q_a=\tilde\rho\,v_a$,
so the momentum flux seen in the cosmic rest frame is simply due to matter moving past those observers. By contrast with the kinematic relations above, here we are decomposing the {\em same} physical tensor with respect to two different congruences, so only $v_a$ is needed to relate the fluid variables.
For the dust model considered here, the matter frame is geodesic and comoving with the fluid, while in the chosen cosmic rest frame, the only nontrivial matter variable is the momentum flux generated by the relative velocity: 
\bea \label{mf}
\mbox{\bf matter frame}~\tilde u^a:~&&
\tilde A_a=0\,,\qquad
\tilde p=0\,,\qquad
\tilde q_a=0\,,\qquad
\tilde\pi_{ab}=0\,,
\\
\label{crf1}
\mbox{\bf cosmic rest frame}~u^a:~&&
\rho=\tilde\rho\,,\qquad
p=0\,,\qquad
q_a=\rho v_a\,,\qquad
\pi_{ab}=0\,.
\eea
For a general cosmic rest frame  the shear, vorticity and acceleration could all be non-zero. In the choice of frame made by \tsagas , we have
\be
A_a\neq0\quad\text{and}\quad\sigma_{ab}=0=\omega_{ab}\,.
\ee

\subsection{Field equations}

The evolution and constraint equations of the covariant approach follow from the Ricci identity, the Bianchi identities, Einstein's equations, and energy-momentum conservation.  For the present work we need only the  equations below~-- we give these for a general spacetime.

The Raychaudhuri equation is
\bea
\label{raych}
\dot\Theta
= -\frac13\Theta^2 + \ud_aA^a - 4\pi G\rho + A_aA^a - \sigma_{ab}\sigma^{ab} + \omega_{ab}\omega^{ab}\,.
\eea
The shear propagation equation is
\bea
\label{shea}
\dot\sigma_{\langle ab\rangle}
= -\frac23\Theta\sigma_{ab}
+\ud_{\langle a}A_{b\rangle}
+4\pi G\pi_{ab}
-E_{ab}
-\sigma_{c\langle a}\sigma_{b\rangle}^{\ c}
+A_{\langle a}A_{b\rangle}
+\omega_{c\langle a}\omega_{b\rangle}^{\ c}\,,
\eea
where $E_{ab} {=C_{acbd}u^c u^d}$ is the electric Weyl tensor, whose propagation equation is
\bea
\label{ew}
\dot E_{\langle ab\rangle} &=&
-\Theta E_{ab}
+{\rm curl}\,H_{ab}
-4\pi G(\rho+p)\sigma_{ab}
-4\pi G\,\ud_{\langle a}q_{b\rangle}
-4\pi G\,\dot\pi_{\langle ab\rangle}
-\frac{4\pi G}{3}\pi_{ab}
\nonumber\\
&&
-A_{\langle a}q_{b\rangle}
+2A^c\varepsilon_{cd(a}H_{b)}^{\ d}
+3\sigma_{c\langle a}E_{b\rangle}^{\ c}
-\frac12\sigma_{c\langle a}\pi_{b\rangle}^{\ c}
-\omega_{c(a}\big(E_{b)}^{\ c}+\pi_{b)}^{\ c}\big)\,.
\eea
Here {$\varepsilon_{abc}$ is the projected alternating tensor, $H_{ab}=\varepsilon_{acd}C^{cd}_{~~be}u^e/2$ is the magnetic Weyl tensor and} the curl is given by
\begin{align}
{\rm curl}\,J_{ab}=\varepsilon_{cd(a}\ud^c J_{b)}^{\ d}\,.
\end{align}
The shear curl constraint is
\bea
\label{hab}
{\rm curl}\,\sigma_{ab}
= H_{ab}
-\frac12\varepsilon_{cd\langle a}\big(\ud_{b\rangle}-2A_{b\rangle}\big)\omega^{cd}\,.
\eea

Finally, energy-momentum conservation gives
\bea
\label{enc}
\dot\rho &=& -(\rho+p)\Theta - \ud^a q_a - 2A^a q_a - \sigma^{ab}\pi_{ab}\,,
\\
\label{momc}
h_a^{\ b}\dot q_b &=& -\frac43\Theta q_a-(\rho+p)A_a-\ud_a p-\ud^b\pi_{ab}-A^b\pi_{ab}-(\sigma_a^{\ b}+\omega_a^{\ b})q_b\,.
\eea
These equations simplify considerably in the dust and linearised settings considered below, where $\pi_{ab}=0$ and nonlinear products of perturbations are neglected.

\section{Peculiar velocities: standard or stronger growth?}
\label{secpec}

Peculiar velocities are not uniquely defined until a reference congruence has been chosen. In metric-based perturbation theory this choice is encoded in the gauge; in the covariant approach it is encoded in the choice of 4-velocity $u^a$. At first order, the two descriptions are equivalent \cite{Dunsby:1992,Ellis:2012}. Thus any comparison between covariant and metric-based peculiar velocities must first ensure that the same reference congruence is being used.

In the standard relativistic treatment of scalar perturbations of FLRW, peculiar velocities are often described in longitudinal gauge. The observer congruence $u^a$ is then the unit normal to the $t=\,$const hypersurfaces, and the matter 4-velocity is $\tilde u^a=u^a+v^a$ to first order. The peculiar velocity $v^a$ is therefore the matter velocity measured relative to that observer congruence. As discussed in the Appendix, this is the covariant counterpart of the physical peculiar velocity in longitudinal gauge. By contrast, in matter-comoving gauges the peculiar velocity vanishes identically. Peculiar velocities are therefore gauge- or frame-dependent, but this is not problematic, provided that physical observables are treated consistently.

The analyses in \tsagas\ consider models close to Einstein-de Sitter and use a covariant formulation based on \cite{Maartens:1998qw,Ellis:2001ms}. For a direct comparison with the standard longitudinal-gauge result, we restrict attention to scalar first-order perturbations. In particular, the matter vorticity vanishes,
\begin{align}
\label{omf}
\tilde\omega_{ab}=0\,.
\end{align}
In longitudinal gauge, the observer congruence normal to the constant-$t$ hypersurfaces is shear-free and irrotational \citep{Malik:2008im}:
\begin{align}
\label{sigf}
\sigma_{ab}=0\,,\qquad \omega_{ab}=0\,.
\end{align}
These conditions do not impose any physical restriction on the matter. They simply characterise the chosen reference congruence. It is therefore natural to make the same choice in the covariant description, as in \cite{Maartens:1998qw,Ellis:2001ms}. With this identification, the covariant setup is the first-order counterpart of longitudinal gauge.

\subsection{Covariant analysis recovers the standard result}
\label{secstand}

With the choice \eqref{sigf}, the covariant equations determine the pair $(v_a,A_a)$ as the degrees of freedom of the frame. The first relation follows immediately from momentum conservation in the $u^a$ frame. Using
$q_a=\rho v_a$, $p=\pi_{ab}=0$, and neglecting products of first-order quantities, \eqref{momc} gives
\begin{align}
\dot q_a+\frac43\Theta q_a+\rho A_a=0\,,
\end{align}
so that
\begin{align}
\label{crf1b}
A_a=-\dot v_a-\frac13\Theta v_a=-\dot v_a-Hv_a\,.
\end{align}

The second relation follows from the shear and Weyl evolution equations. Since the reference congruence is shear-free, the linearised shear propagation equation \eqref{shea} reduces to
\begin{align}
\label{ewac}
E_{ab}=\ud_{\langle a}A_{b\rangle}\,.
\end{align}
Similarly, since the congruence is irrotational, the shear-curl constraint gives
\begin{align}
H_{ab}=0\,.
\end{align}
Substituting these results into the linearised electric Weyl propagation equation \eqref{ew}, and using $q_a=\rho v_a$, yields
\begin{align}
\label{dad}
\big(\ud_{\langle a}A_{b\rangle}\big)^{\displaystyle\cdot}
=
-3H\,\ud_{\langle a}A_{b\rangle}
-4\pi G\rho\,\ud_{\langle a}v_{b\rangle}\,.
\end{align}
Using the linearised commutation relation
\begin{align}
\label{comm}
\big(\ud_a J_b\big)^{\displaystyle\cdot}=\ud_a\dot J_b-H\,\ud_a J_b\,,
\end{align}
valid for the scalar, shear-free and irrotational setting considered here \citep{Maartens:1998qw,Ellis:2001ms}, this becomes
\begin{align}
\label{dad2}
\ud_{\langle a}\Big[\dot A_{b\rangle}+2H A_{b\rangle}+4\pi G\rho\,v_{b\rangle}\Big]=0\,.
\end{align}
For scalar perturbations, the bracketed quantity is itself a projected gradient. Hence vanishing projected, symmetric, trace-free derivative implies that, up to a spatially homogeneous mode which is not part of the perturbations of interest, the bracket vanishes:
\begin{align}
\label{dad3}
\dot A_a+2H A_a+4\pi G\rho\,v_a=0\,.
\end{align}

Equations \eqref{crf1b} and \eqref{dad3} form a closed linear system for $(v_a,A_a)$. Eliminating $A_a$ gives the key equation for the peculiar velocity field:
\begin{align}
\label{pvst}
\ddot v_a+3H\dot v_a+\big(\dot H+2H^2-4\pi G\rho\big)v_a=0\,.
\end{align}
In Einstein-de Sitter,
\begin{align}
\dot H=-\frac32H^2\,,\qquad 4\pi G\rho=\frac32H^2\,,
\end{align}
and therefore
\begin{align}
\ddot v_a+3H\dot v_a-H^2v_a=0\,.
\end{align}
The growing solution is
\begin{align}
\label{pvst1}
v_a\propto t^{1/3}\propto a^{1/2}\,,
\end{align}
in agreement with the standard longitudinal-gauge result derived in the Appendix.

Equation \eqref{crf1b} then gives the corresponding behaviour of the acceleration of the chosen reference congruence:
\begin{align}
A_a\propto t^{-2/3}\propto a^{-1}\,.
\end{align}
Thus the acceleration required to keep the cosmic rest frame at rest relative to the matter decays with time, as expected in a decelerating dust universe.

The remaining scalar perturbations are determined once $(v_a,A_a)$ are known. In the covariant approach we define the 
density and expansion gradients as
\begin{align}
\Delta_a=\frac{a}{\rho}\ud_a\rho\,,\qquad Z_a=a\,\ud_a\Theta\,.
\end{align}
Linearising the energy conservation equation and taking a projected gradient gives
\begin{align}
\label{DZsys1}
\dot\Delta_a=-Z_a-a\,\ud_a(\ud^b v_b)-3aH\,A_a\,,
\end{align}
while taking a projected gradient of the Raychaudhuri equation gives
\begin{align}
\label{DZsys2}
\dot Z_a=-2H\,Z_a-\frac32H^2\Delta_a-\frac92 aH^2A_a+a\,\ud_a\ud^bA_b\,.
\end{align}
Hence $(\Delta_a,Z_a)$ form a forced linear subsystem once $(v_a,A_a)$ are known. In particular, the source terms in \eqref{DZsys1}--\eqref{DZsys2} are fixed by the already-determined velocity and acceleration fields. The density and expansion gradients therefore follow from the closed $(v_a,A_a)$ sector; they do not generate independent growth laws for $v_a$.

It should be noted that although the shear-free frame choice is often referred to as `quasi-Newtonian', and the corresponding gauge choice the `Newtonian gauge', the solution remains a fully relativistic treatment of linear perturbations of EdS~-- it is \emph{not} a Newtonian approximation.

\subsection{Ruling out faster growth}
\label{secstrong}

The origin of the spurious stronger growth laws claimed in \tsagas\ can now be isolated. The covariant system may be rearranged in several equivalent ways to give second-order differential equations for $v_a$ such that they appear to have a source driving them. Two examples are especially relevant.
The first is Eq.~(15) of \cite{Maglara:2022pmz}, specialised to Einstein-de Sitter:
\begin{align}
\label{eq:eds_eq15}
\ddot v_a+\frac12H\dot v_a-2H^2 v_a
=
\frac{1}{3aH}\Big[a\,\ud_a(\ud^b v_b)^{\displaystyle\cdot}+\dot Z_a+\ddot\Delta_a\Big]\,.
\end{align}
The second is Eq.~(11) of \cite{Tsagas:2024zrx}:
\begin{align}
\label{eq:eds_eq11}
\ddot v_a+H\dot v_a-\frac32H^2 v_a
=
\frac{1}{3aH}\Big[\ddot\Delta_a+2H\dot\Delta_a-\frac32H^2\Delta_a\Big]\,.
\end{align}
Both equations are correct. They are simply different rearrangements of the same underlying linearised covariant system, and direct substitution of \eqref{DZsys1} and \eqref{DZsys2} confirms this.

The error arises when the right-hand side of either \eqref{eq:eds_eq15} or \eqref{eq:eds_eq11} is treated as if it were an external source that may be switched off in order to isolate a physically relevant `homogeneous' solution. This is not a legitimate approximation, because the terms on the right-hand side are not independent of $v_a$. They are fixed by the coupled perturbation equations.

We can see the problem directly by the fact that the same truncation applied to the two equally valid equations above leads to two different growth laws. Setting the right-hand side of \eqref{eq:eds_eq15} to zero gives
\begin{align}
\label{WrongHom15}
\ddot v_a+\frac12H\dot v_a-2H^2v_a=0
\qquad\Longrightarrow\qquad
v_a\propto t^{4/3}\propto a^2\,,
\end{align}
whereas setting the right-hand side of \eqref{eq:eds_eq11} to zero gives
\begin{align}
\label{WrongHom11}
\ddot v_a+H\dot v_a-\frac32H^2v_a=0
\qquad\Longrightarrow\qquad
v_a\propto t\propto a^{3/2}\,.
\end{align}
Since both equations come from the same covariant system, these incompatible `homogeneous' growth laws already show that the truncation of the right hand side of either has no invariant physical meaning. In fact neither growth law is allowed, because both contradict the closed subsystem result \eqref{pvst}.\footnote{
More generally,  one may form the linear combination
$\alpha\times\eqref{eq:eds_eq15}+(1-\alpha)\times\eqref{eq:eds_eq11}$\,,
for any constant $\alpha$,
to obtain
\begin{align*}
\ddot v_a+\Big(1-\frac{\alpha}{2}\Big)H\dot v_a-\Big(\frac32+\frac{\alpha}{2}\Big)H^2 v_a
=
\frac{1}{3aH}\Big[\ddot\Delta_a+(2-2\alpha)H\dot\Delta_a
-(1-\alpha)\frac32H^2\Delta_a
+\alpha\big(a\,\ud_a(\ud^bv_b)^{\displaystyle\cdot}+\dot Z_a\big)\Big]\,.
\end{align*}
The split into a `homogeneous' part for $v_a$ and a `source' term is clearly not unique. If we look for a homogeneous solution without the right-hand side, the resulting growth law would  arbitrarily depend on the choice of $\alpha$.} 

We can see this more generally. The linearised perturbations form a coupled system for $(v_a,A_a,\Delta_a,Z_a)$. The pair $(v_a,A_a)$ is a closed subset of that system: \eqref{crf1b} and \eqref{dad3} determine its time evolution uniquely up to initial data, and give the standard growing mode \eqref{pvst1}. The remaining equations \eqref{DZsys1}--\eqref{DZsys2} then determine $(\Delta_a,Z_a)$ once $(v_a,A_a)$ are known. \emph{They cannot consistently be used to infer additional growth laws for $v_a$, because their source terms depend on the already-determined velocity and acceleration fields.}

Equivalently, \eqref{DZsys1}--\eqref{DZsys2} are a forced linear system for $(\Delta_a,Z_a)$, with forcing proportional to $\ud_a(\ud^b v_b)$ and $\ud_a\ud^bA_b$. Any manipulation of these equations that appears to yield an alternative second-order evolution equation for $v_a$ necessarily treats those forcing terms as if they were independent or negligible, thereby enlarging the solution space beyond that allowed by the full covariant system. This is the origin of the stronger growth laws claimed in \tsagas.

\subsection{If the chosen congruence is not shear-free}
\label{secshear}

If the shear-free condition $\sigma_{ab}=0$ is relaxed, the 
$(v_a,A_a)$ system in Sec. \ref{secstand} is no longer closed.
(We restrict to the scalar sector, so that $\omega_{ab}$ and the magnetic Weyl tensor carry no independent scalar degree of freedom at first order.)  The full linearised
covariant system for scalar dust perturbations then comprises the five
evolution equations\footnote{From the equations of Sec. \ref{seccov}:
\eqref{vdot} is momentum conservation \eqref{momc}; \eqref{sigmadot} is
shear evolution \eqref{shea}; \eqref{Edot} is electric Weyl evolution
\eqref{ew}; \eqref{Deltadot} and \eqref{Zdot} follow from energy
conservation \eqref{enc} and the Raychaudhuri equation \eqref{raych} by
taking projected gradients, using the linear commutator
$(\ud_a f)^{\displaystyle\cdot}=\ud_a\dot f-H\,\ud_a f-A_a\dot f$ for
scalars, where the $A_a\dot f$ piece sources the $A_a$ terms in the
latter two.}
\begin{align}
\label{vdot}
\dot v_a + Hv_a &= -A_a\,,\\
\label{sigmadot}
\dot\sigma_{ab}+2H\sigma_{ab}+E_{ab} &= \ud_{\langle a}A_{b\rangle}\,,\\
\label{Edot}
\dot E_{ab}+3HE_{ab}+4\pi G\rho\,\sigma_{ab}  +4\pi G\rho\,\ud_{\langle a}v_{b\rangle}&=0\,,\\
\label{Deltadot}
\dot\Delta_a+Z_a+a\,\ud_a(\ud^b v_b) &= -3aH\,A_a\,,\\
\label{Zdot}
\dot Z_a+2HZ_a+\tfrac32H^2\Delta_a &= -\tfrac92 aH^2\,A_a+a\,\ud_a\ud^b A_b\,.
\end{align}
After scalar harmonic decomposition this is a evolution system for the 5 scalar
amplitudes associated with
\be
(v_a,\sigma_{ab},E_{ab},\Delta_a,Z_a).
\ee
\emph{It contains no evolution equation for the sixth variable, the acceleration amplitude.}
 The 4-acceleration $A_a$
appears as a source on the right-hand side of every equation except
\eqref{Edot}~-- yet \emph{it has no evolution equation of its own}. Energy conservation appears in this system as \eqref{Deltadot}, the
propagation equation for the comoving density gradient; the
`relativistic 4-acceleration formula' of \cite{Tsaprazi:2019bbi} is just \eqref{Deltadot} rearranged. 
The system is one equation short and does not close: the time evolution of
$A_a$ is not fixed by the dust dynamics. This is simply the frame degree of freedom inherent in the 1+3 covariant approach, specialised to perturbations of EdS. This is essentially a gauge choice (though formally slightly different), and behaviour associated with it is akin to a gauge mode.  

The missing equation must therefore be supplied by a frame/gauge choice
on $A_a$. The shear-free condition $\sigma_{ab}=0$ used in
Sec. \ref{secstand} does exactly this: \eqref{sigmadot} reduces to the
constraint $E_{ab}=\ud_{\langle a}A_{b\rangle}$, and substitution into
\eqref{Edot} fixes the evolution of $A_a$~-- giving \eqref{dad3}~-- and
hence the standard growth law $v_a\propto t^{1/3}$. This is just part of the normal gauge fixing of perturbation theory. The shear-free frame is of interest as it is the covariant counterpart of the time slicing of the Newtonian/longitudinal gauge~-- \emph{it is not a physical restriction on the solution space}. One may choose a hypersurface-orthogonal reference congruence
$u_a=-N\nabla_a t$, with lapse chosen so that
$A_a=\ud_a\ln N$ has any prescribed first-order time dependence, for example
$A_a\propto t^n$. The resulting growth of the shear of this congruence and of
the dust velocity measured relative to it is a property of the chosen
frame/slicing, {\em not an additional dynamical mode of the dust.} 

Other problems with \tsagas\ are also obvious in second-order form. Eliminating $E_{ab}$
between \eqref{sigmadot} and \eqref{Edot} and using \eqref{vdot} gives the
coupled velocity-shear equation in Einstein-de Sitter,
\begin{align}
\label{shear_main}
\ddot\sigma_{ab}+5H\dot\sigma_{ab}+4\pi G\rho\,\sigma_{ab}
=-\ud_{\langle a}\big[\ddot v_{b\rangle}+3H\dot v_{b\rangle}-H^2v_{b\rangle}\big]\,.
\end{align}
This is one second-order tensor equation in two tensor unknowns: it
relates $\sigma_{ab}$ to $v_a$ but does not determine either separately.
The same coupled system can equally be rearranged into different
second-order equations for $v_a$ alone~-- including \eqref{eq:eds_eq15}
and \eqref{eq:eds_eq11} of Sec. \ref{secstrong}~-- each with a different
`source' on the right-hand side. Setting that source to zero produces a
different `homogeneous' growth law: \eqref{eq:eds_eq15} gives
$v_a\propto t^{4/3}$, \eqref{eq:eds_eq11} gives $v_a\propto t$, and so on.
\emph{None is selected by the dust dynamics.} The right-hand sides are not
external sources but are constrained by the same coupled system
\eqref{vdot}--\eqref{Zdot}, and which truncation one chooses determines
which spurious `homogeneous' growth one extracts. The rapid growth modes
obtained in \tsagas\ are just these truncation artefacts, and have nothing to do with a `relativistic solution'.\footnote{For a fixed linear operator \(L\), the split \(L[y]=S\) into
homogeneous modes \(L[y]=0\) plus a sourced Green-function integral is
correct only when \(S\) is prescribed independently of \(y\).  Here the right-hand side is not a source in this Green-function
sense: it is a constrained combination of the same unknowns, so
setting it to zero replaces the coupled system by a different system.}

The dust is geodesic, $\tilde A_a=0$, so $A_a$ in some chosen frame is not
a force on the matter but the rate at which the auxiliary congruence is
externally accelerated relative to the freely-falling dust. From
\eqref{vdot}, sustaining any growing $v_a$ therefore requires an additional 
prescription to supply the corresponding $A_a$. The linear law $v_a\propto t$
now favoured in~\cite{Tsagas:2024zrx} requires constant $A_a$ in
Einstein-de Sitter; the steeper $v_a\propto t^{4/3}$ requires
$A_a\propto t^{1/3}\propto a^{1/2}$. No physical agency supplies any such
$A_a$. The only physically motivated non-comoving congruence is the CMB
frame, whose 4-acceleration is fixed by radiation momentum conservation~\citep{Maartens:1998xg},
\begin{align}
A_a=-\frac14\,\frac{\ud_a\rho_\gamma}{\rho_\gamma}
-\frac{3}{4\rho_\gamma}\,\ud^b\pi^{(\gamma)}_{ab}\,,
\end{align}
and decays after decoupling.

The conclusion is starkest in the case $E_{ab}=0$. With $H_{ab}=0$
already imposed, this is the conformally flat case and the spacetime is exact
FLRW~-- there is no physical perturbation at all. Equation \eqref{Edot}
then forces $\sigma_{ab}=-\ud_{\langle a}v_{b\rangle}$, which is just the
kinematic relation $\tilde\sigma_{ab}=0$ for a shear-free matter frame.
Both the velocity field and the corresponding 4-acceleration
$A_a=-\dot v_a-Hv_a$ are arbitrary: any choice of tilted observer
congruence in unperturbed FLRW is admissible. A growth law such as
$v_a\propto t$ or $v_a\propto t^{4/3}$ is then nothing more than one
such choice of tilt; the dust dynamics neither selects nor forbids it,
and without a physical force to drive the corresponding $A_a$ the
growth law is irrelevant.

\section{Discussion and conclusion}
\label{secdisc}

The results of Sec. \ref{secpec} clearly show that the faster peculiar-velocity growth claimed in \tsagas\ is \emph{not} a solution of the full covariant system. The covariant, shear-free and irrotational reference congruence used there leads instead to the standard relativistic growth law in Einstein--de Sitter,
\begin{align}
v_a\propto t^{1/3}\propto a^{1/2}\,,
\end{align}
as given in \eqref{pvst1}. The corresponding acceleration of the chosen cosmic rest frame decays as
$A_a\propto a^{-1}\,,$
following from \eqref{crf1b}. There is therefore no new fast-growing relativistic mode in the covariant approach. Standard perturbation theory captures all linear relativistic effects consistently at first-order.

This resolves the origin of the stronger growth laws claimed in \tsagas. Once the shear-free condition is imposed, the pair $(v_a,A_a)$ forms a closed subsystem: \eqref{crf1b} and \eqref{dad3} determine its evolution uniquely up to initial data, leading to \eqref{pvst} and hence to \eqref{pvst1}. The remaining density and expansion gradients then follow from the forced system \eqref{DZsys1}--\eqref{DZsys2}. They do not supply additional admissible velocity modes. The stronger laws obtained by setting the right-hand side of equations such as \eqref{eq:eds_eq15} or \eqref{eq:eds_eq11} to zero arise because terms constrained by the coupled system are treated as if they were independent sources. Different frame choices are possible such as $A_a=\,$const. with non-zero shear, but these are no more than gauge degrees of freedom with no physical relevance.

These points also clarify the associated claims about bulk flows in \tsagas. Since the faster growth laws \eqref{WrongHom15} and \eqref{WrongHom11} are spurious, they cannot provide an explanation for anomalously large bulk flows. Any tension between observed bulk flows and the standard cosmological model must therefore be addressed on other grounds.

A further claim in \tsagas\ is that bulk motions in a matter-dominated universe can lead observers to infer accelerated expansion, even though the underlying universe is decelerating. {In~\cite{Sah:2024csa} it is claimed that a dipole found in the Pantheon+ SNIa data leads the conclusion that ``The inferred cosmic acceleration cannot therefore
be due to a Cosmological Constant, but is likely a general relativistic effect due
to the anomalous bulk flow in our local Universe.'' This is clearly not supported by the correct peculiar velocity evolution. } Irrespective of the details of the growth law, there is a {futher} problem with the interpretation of different congruences. The matter congruence, with 4-velocity $\tilde u^a$, defines the {\em physical} expansion of the dust and hence of the late-time universe traced by galaxies. By contrast, the chosen `cosmic rest frame' with 4-velocity $u^a$ is an auxiliary observer congruence introduced in order to describe peculiar motions. Its kinematics do {\em not} determine the physical expansion of the matter, {and has no physical meaning}.

This is seen directly from the Raychaudhuri equation for the matter congruence. Writing
\begin{align}
\tilde{\mathsf H}=\frac13\tilde\Theta=\frac13\nabla_a\tilde u^a\,,
\end{align}
the Ricci identity \eqref{raych} gives, in the matter frame for {dust}
\begin{align}
\tilde u^a\nabla_a \tilde\Theta
=
-\frac13\tilde\Theta^2
-4\pi G\tilde\rho
-\tilde\sigma_{ab}\tilde\sigma^{ab}\,,
\label{raym2}
\end{align}
and therefore {the covariant version of the deceleration parameter, as defined in \tsagas, is}
\begin{align}
\tilde{\mathsf q}
:=
-\frac{3}{\tilde\Theta^2}
\left(
\tilde u^a\nabla_a\tilde\Theta+\frac13\tilde\Theta^2
\right)
=
\frac{1}{3\tilde{\mathsf H}^2}
\left(
4\pi G\tilde\rho+\tilde\sigma_{ab}\tilde\sigma^{ab}
\right)
>0\,.
\label{tsq2}
\end{align}
Thus the matter-dominated universe is covariantly decelerating. This physical deceleration is a property of the matter worldlines and is independent of the auxiliary congruence used to define peculiar velocities.

The quantity denoted $\mathsf q$ for the chosen cosmic rest frame describes the deceleration of the $u^a$ congruence itself. It is not the same as the deceleration of the matter congruence as inferred observationally by observers moving with respect to the matter. Relating observations made in different states of motion requires the appropriate boost and lightcone relations. It is therefore meaningless to compare $\tilde{\mathsf q}$ and $\mathsf q$ as if they referred to the same physical quantity.

The same point applies to claims of a dipole in the deceleration tensor generated by bulk flows. Any dipole in the deceleration of galaxy worldlines should be defined for the matter congruence and then transformed to the frame of the observer.
It cannot be inferred directly from differences between tensors defined on different congruences. In particular, any observed dipole in a deceleration parameter must be analysed through the covariant cosmographic expansion and its transformation under boosts \citep{Heinesen:2020bej,Maartens:2023tib,Adamek:2024hme}, rather than by comparing the kinematics of two different congruences.

In summary, the covariant approach does not replace the standard relativistic treatment of peculiar velocities. When applied consistently, it reproduces it. The stronger growth laws claimed in \tsagas\ arise from an inconsistent truncation of the coupled covariant system, and the associated claims about bulk flows and apparent acceleration do not follow. \\

\noindent\textbf{Note added in v2.} A response, \cite{Tsagas:2026XX}, appeared
after Version 1 of this paper, in which the present authors are  inducted
into a confederacy~\citep{Swift1727}. Swift indeed warned that, ``When a true genius appears in the world, you may know him by this sign,
that the dunces are all in confederacy against him.''  The points
\cite{Tsagas:2026XX} raises led these dunces  to add Sec. \ref{secshear}, with our conclusion  unchanged.

The central issue is simple. The shear-free reference
congruence used above is a frame choice: it is the covariant counterpart of
longitudinal gauge in metric perturbation theory. It is not a Newtonian
approximation. If this frame condition is removed, the dust equations do not
determine the 4-acceleration of the auxiliary congruence. The rapid growth laws
then correspond to prescribed accelerations of that congruence, not to
additional physical modes of the dust. In EdS, a tilted frame with $v_a\propto t$ requires
constant $A_a$, while $v_a\propto t^{4/3}$ requires
$A_a\propto t^{1/3}$.  These are perfectly valid frame choices from a mathematical point of view: we are  at liberty to discuss a frame with $A_a=\,$const., but the \emph{only} physical frames are that of the dust (which has $A_a=0$) and the CMB (in which $A_a$ is decaying). \emph{No physical component in the model supplies these acceleration laws.}

A specific claim in \cite{Tsagas:2026XX} is that the shear-free analysis
omits energy conservation, and with it the `gravitational contribution
of the peculiar flux'. This is a misreading of the structure of the equations, not to mention our v1. Energy
conservation is part of the linearised covariant system throughout this
paper: it appears in Sec. \ref{secshear} as \eqref{Deltadot}, the
propagation equation for the comoving density gradient.  The peculiar-flux contribution to gravity~-- $q_a=\rho
v_a$ entering momentum conservation through $\rho A_a$ and shear and
Weyl propagation through $\ud_{\langle a}v_{b\rangle}$~-- is fully
present in both formulations.

~\\
\noindent{\bf Acknowledgements:} We thank Timothy Clifton,  Asta Heinesen, Pedro Ferreira, Mohamed Rameez and Subir Sarkar for comments and discussions.
CC is supported by STFC grant ST/X000931/1. 
RM was supported by the South African Radio Astronomy Observatory and the National Research Foundation (grant no. 75415). 

\appendix

\section{Peculiar velocities in perturbed FLRW}\label{appb}

Considering peculiar velocities in perturbed FLRW (Einstein-de Sitter), in order to connect with the covariant discussion it is useful to distinguish carefully between  the 4-velocity of the longitudinal-gauge observers $u^a$ (i.e. the unit normal to the constant-$t$ hypersurfaces), the matter 4-velocity $\tilde u^a$, the physical peculiar 3-velocity $v^a$ and the coordinate
velocity $\sd x^i/\sd t$.

Scalar perturbations of Einstein-de Sitter  in Newtonian (longitudinal) gauge, with vanishing
anisotropic stress, are described by
\begin{equation}\label{eq:NGmetric_appB}
\sd s^2=-(1+2\Phi)\,\sd t^2+a^2(t)\,(1-2\Phi)\,\delta_{ij}\,\sd x^i \sd x^j \,.
\end{equation}
The Newtonian-gauge observer worldlines are normal to the $t=\,$const hypersurfaces. Their
4-velocity is therefore
\begin{equation}\label{eq:u_appB}
u^\mu=\big(1-\Phi,0,0,0\big), \qquad
u_\mu=\big(-(1+\Phi),0,0,0\big),
\end{equation}
to first order.
The matter worldlines are described by a different 4-velocity $\tilde u^\mu$. Writing
\begin{equation}
\tilde u^\mu=\frac{\sd x^\mu}{\sd\tau}
=\Big(\frac{\sd t}{\sd\tau},\frac{\sd x^i}{\sd\tau}\Big)
=\frac{\sd t}{\sd\tau}\Big(1,\frac{\sd x^i}{\sd t}\Big),
\end{equation}
the normalisation condition $\tilde u^\mu \tilde u_\mu=-1$ gives, 
\begin{equation}
\tilde u^\mu=\Big(1-\Phi,\frac{\sd x^i}{\sd t}\Big),
\end{equation}
if one wishes to use the coordinate velocity $\sd x^i/\sd t$.
In the covariant decomposition one defines the peculiar velocity by 
\begin{equation}\label{eq:boost_appB}
\tilde u^a=\gamma\big(u^a+v^a\big), \qquad u_a v^a=0.
\end{equation}
Since $v^a$ is orthogonal to $u^a$, it is a physical 3-velocity measured by the Newtonian-gauge
observers. 
Using \eqref{eq:u_appB}, this means
\begin{equation}
v^\mu=\tilde u^\mu-u^\mu=\Big(0,\frac{v^i}{a} \Big),
\end{equation}
where $v^i$ is the physical peculiar velocity. Therefore
\begin{equation}\label{eq:utilde_phys_appB}
\tilde u^\mu=\Big(1-\Phi,\frac{v^i}{a}\Big),
\qquad
\tilde u_\mu=\big(-(1+\Phi),a\,v_i\big),
\end{equation}
where $v_i=\delta_{ij}v^j$.
Therefore the physical
peculiar velocity and the coordinate velocity are related by
\begin{equation}\label{eq:vcoordrel_appB}
v^i=a\,\frac{\sd x^i}{\sd t}.
\end{equation}
Thus the covariant peculiar velocity $v^a$ is associated with the physical peculiar velocity $v^i$,
not with the coordinate velocity $\sd x^i/\sd t$.

For dust, $\nabla_\mu T^{\mu\nu}=0$ gives the linear Euler equation in Newtonian gauge,
\begin{equation}\label{eq:Euler_appB}
\dot v_i + H v_i = -\frac{1}{a}\,\partial_i\Phi\,.
\end{equation}
The linearised $0i$ Einstein equation gives
\begin{equation}\label{eq:0i_appB}
\partial_i\big(\dot\Phi+H\Phi\big)= -4\pi G\,a\,\rho\,v_i\,,
\end{equation}
where $\rho$ is the background matter density.

Differentiating \eqref{eq:Euler_appB}, we find
\begin{equation}
\ddot v_i+\dot H\,v_i+H\dot v_i
= -\frac{1}{a}\partial_i\dot\Phi+\frac{H}{a}\partial_i\Phi\,.
\end{equation}
Using \eqref{eq:0i_appB}, this becomes
\begin{equation}
\ddot v_i+\dot H\,v_i+H\dot v_i
= \frac{2H}{a}\partial_i\Phi+4\pi G\rho\,v_i\,.
\end{equation}
Eliminating $\partial_i\Phi$ via \eqref{eq:Euler_appB}, 
we obtain
\begin{equation}\label{eq:vi_eq_appB}
\ddot v_i+3H\dot v_i+\big(\dot H+2H^2-4\pi G\rho\big)v_i=0\,.
\end{equation}
Using
\begin{equation}
\dot H=-\frac32H^2,
\qquad
4\pi G\rho=\frac32H^2,
\end{equation}
it follows that
\begin{equation}\label{eq:vi_EdS_appB}
\ddot v_i+3H\dot v_i-H^2v_i=0,
\end{equation}
which is \eqref{pvst}. Therefore the growing mode is
\begin{equation}
v_i\propto t^{1/3}\propto a^{1/2}.
\end{equation}

By contrast, the coordinate velocity
\begin{equation}
w^i=\frac{\sd x^i}{\sd t}=\frac{v^i}{a}
\end{equation}
obeys
\begin{equation}
\dot w_i+2H w_i=-\frac{1}{a^2}\partial_i\Phi,
\end{equation}
and therefore has the dominant mode
\begin{equation}
w_i\propto a^{-1/2}\propto t^{-1/3}.
\end{equation}

\bibliographystyle{apsrev4-1}
\bibliography{OJA-v1}

\end{document}